\documentstyle[12pt]{article}

\title{Synergetic theory for jamming transition \\ in traffic flow}

\author{ Alexander~I.~Olemskoi, Alexei~V.~Khomenko \\
{\it Sumy State University} \\ {\it  2, Rimskii-Korsakov St.,
40007 Sumy, Ukraine} \\
{\it E-mail: olemskoi@ssu.sumy.ua }}

\date{\ }
\begin{document}
\maketitle{
\vspace*{-1.8cm}

\thispagestyle{empty}
\begin{abstract}
The theory of a jamming transition is proposed for the homogeneous
car-following model within the framework of Lorenz scheme. We
represent a jamming transition as a result of the spontaneous deviations
of headway and velocity that is caused by the acceleration/braking
rate to be higher than the critical value. The
stationary values of headway and velocity deviations, and time of
acceleration/braking are derived as functions of control
parameter (time needed for car to take the characteristic velocity).

\vspace{0.4cm}

\noindent PACS number(s): 05.70.Fh, 05.70.Jk, 89.40.+k


\end{abstract}
}
\section{Introduction}

In recent years considerable study has been given to the traffic
problems \cite{1a}. It is shown, in particular,  that the jamming
transition is similar to the conventional gas-liquid phase
transition, where the freely moving traffic and the jammed traffic
correspond to the gas and liquid phases respectively. The transition
between them is caused by growth of car density above a critical
value. The congested traffic flow with unstable uniform part leads to
the formation of traffic jams where the freely moving traffic and
jammed traffic coexists. Within the framework of Ref.~\cite{2a},
the jamming transition is represented as a first-order phase transition,
whose behavior is defined by headway (car density)
that acts as the volume (density) and by the inverted
delay time (sensitivity parameter) that reduces to temperature.

Our approach is to take into consideration the complete set of
freedom degrees as equivalent variables. We obtain the
self-consistent analytical description of the jamming transition
as result of the self-organization caused by the positive feedback
of the headway deviation and acceleration/braking time
-- on the one hand, as well as the negative feedback
of the deviations of headway and velocity -- on the other one.

The paper is organized as follows. In Section 2
the self-consistent Lorenz system of the governed equations
for the headway and velocity deviations as well as for
the acceleration/braking time is obtained.
The jamming transition is shown to be supercritical
in character (has the second order)
if a relaxation time for the first of pointed out quantities
does not depend on its value; it transforms to subcritical regime
with this dependence appearance.
Section 3 deals with the determination of steady-state values
for the headway deviation and the acceleration/braking time
within adiabatic approximation. Out of the latter limit,
the time dependencies for the headway and velocity deviations
are studied on the basis of the phase-portrait method.
Section 4 contains a short discussion of used assumptions.

\section{Basic equations}

Within framework of the simplest car-following model the acceleration
$\dot V$ of a given vehicle as a function of its distance $\Delta x$
to the front vehicle is defined by equality $\dot V {=}
[v_{opt}(\Delta x)-V]/\tau$, where $v_{opt}(\Delta x) = \Delta x/t_0$
is the optimal velocity function ($t_0$ being a characteristic time
interval), $h = Vt_0$ is the optimal headway and $\tau$ is the
time of acceleration/braking needed for car to take the optimal velocity.
It is convenient to introduce deviations $\eta\equiv \Delta x-h$
and $v\equiv \Delta\dot x-h/t_0+V$ of
headway $\Delta x$ and its velocity $\Delta\dot x$ from corresponding
optimal values $h$ and $h/t_0-V$. Then, the flow of cars can be described
in terms of the pointed out quantities $\eta$, $v$, and $\tau$.
The key point of our approach is that the above degrees of freedom
are assumed to be of dissipative type, so that, when they are not coupled,
their relaxation to the steady state is governed by the Debye-type
equations with corresponding relaxation times
$t_{\eta}$, $t_{v}$, $t_{\tau}$. Within the simplest approach, equations for
the time dependencies $\eta (t)$, $v(t)$, and $\tau(t)$ are supposed to
coincide formally with the Lorenz system that describes the
self-organization process \cite{3a}.

The first of the stated equations has the form
\begin{equation}
\dot\eta= -\eta / t_{\eta} + v, \label{2}
\end{equation}
where the dot stands for a derivative with respect to time $t$. The
first term in the right-hand side describes the Debye relaxation
during time $t_\eta$, the second one is the usual addition.
In a stationary state, when $\dot\eta=0$, solution
of Eq.~(\ref{2}) defines conventional linear relationship $\eta=t_\eta v$, so
that the headway deviation is proportional to the velocity deviation.

The equation for the rate of quantity $v$ variation
is supposed to have the nonlinear form
\begin{equation} \dot v = - v /t_v + g_v \eta \tau , \label{3}
\end{equation}
where $t_v,~g_v$ are positive constants. As in Eq.~(\ref{2}), the
first term in the right-hand side of Eq.~(\ref{3}) describes the
relaxation process of velocity deviation $v$ to the stationary value
$v = 0$ determined by a time $t_v$. The second term describes
the positive feedback of the headway deviation $\eta$ and the
time $\tau$ of acceleration/braking on the velocity deviation $v$ that
results in the increase of value $v$ and, thus, causes the
self-organization process.

The kinetic equation for the acceleration/braking time $\tau$
\begin{equation} \dot \tau = (\tau_0-\tau)/t_\tau - g_\tau \eta v \label{6}
\end{equation}
differs from Eqs.~(\ref{2}), (\ref{3}) as
follows: the relaxation of quantity $\tau$ occurs not to the zero but
to the finite value $\tau_0$, representing the stationary
time needed for car to take the characteristic velocity
(in other words, $\tau_0$ is the car characteristic);
$t_\tau$ is a corresponding relaxation time.
In Eq.~(\ref{6}) the negative feedback
of the quantities $\eta$ and $v$ on $\tau$ is introduced to imply the
decrease of acceleration/braking time $\tau$ with the growth
of the headway and velocity deviations ($g_\tau>0$ is a corresponding
constant).

The equations (\ref{2}), (\ref{3}), (\ref{6}) constitute the basis for
self-consistent description of the car-following model
driven by the control parameter $\tau_0$. The distinguishing
feature of these equations is that nonlinear terms that enter
Eqs.~(\ref{3}), (\ref{6}) are of opposite signs, while Eq.~(\ref{2}) is
linear.  Physically, the latter means just that the velocity deviation is
the derivative of headway deviation with respect to time. The
negative sign of the last term in Eq.~(\ref{6}) can be regarded as a
manifestation of Le Chatelier principle, i.e. since an decrease in
the acceleration/braking time
promote to the formation of a stable car
flow, the headway and velocity deviations $\eta$ and $v$
tend to impede the growth of the acceleration/braking time
and, as a consequence, the jamming.
The positive feedback of $\eta$ and $\tau$ on $v$ in Eq.~(\ref{3})
plays an important part in the problem. As we will see later, it is
precisely the reason behind the self-organization that brings about
the traffic jam.

To explain the relaxation transition to the stable jamming state,
we will show further that it is quite enough to use the
adiabatic approximation: $t_v=0, t_{\tau}=0$. Therefore we could
proceed not from Eqs.~(\ref{3}), (\ref{6}) but from much simple
expressions
\begin{equation}
v = a_v \eta \tau,\quad a_v\equiv t_v g_v;\qquad
\tau = \tau_0 -a_\tau\eta v,\quad a_\tau\equiv t_\tau g_\tau,
\label{a}
\end{equation}
which are related to the stationary case $\dot v =
0,~ \dot \tau = 0$ in Eqs.~(\ref{3}), (\ref{6}) respectively.
The equalities (\ref{a}) have absolutely clear physical meaning:
the increase of the
headway deviation $\eta$ or acceleration/braking time $\tau$ leads to
growth of the velocity deviation $v$, whereas the increase of the
headway $\eta$ and velocity $v$ deviations should cause the decrease of
acceleration/braking time $\tau$ in comparison with characteristic
time $\tau_0$ if the car flow is not broken.

After introducing the suitable scales for quantities $\eta ,~ v,~ \tau$:
\begin{equation} \eta_m \equiv (a_v a_\tau )^{-1/2},\,
v_m \equiv \eta_m / t_\eta = t_\eta^{-1}(a_v a_\tau )^{-1/2},\,
\tau_c\equiv (t_\eta a_v)^{-1}, \label{7} \end{equation}
Eqs.~(\ref{2}), (\ref{3}), (\ref{6}) can be rewritten in the simplest
form of the well--known Lorenz system: \begin{eqnarray}
\dot{\eta}&=&-\eta + v, \label{8a}\\
\epsilon~\dot{v}&=&-v + \eta \tau, \label{8b}\\
\delta~\dot{\tau}&=&(\tau_{0}-\tau)-\eta v,
\label{8c}
\end{eqnarray}
where the relaxation times ratios $\epsilon\equiv t_{v}/ t_{\eta},\,
\delta\equiv t_{\tau}/ t_{\eta}$ are introduced
and the dot now stands for the derivative with respect to the
dimensionless time $t/\tau_{\eta}$. In general, the system (\ref{8a})
-- (\ref{8c}) can not be solved analytically, but in the simplest case
$\epsilon\ll 1\, {\rm and}\, \delta\ll 1$,
the left-hand sides of Eqs.~(\ref{8b}), (\ref{8c}) can be neglected.
Then, the adiabatic approximation can be used to express
the velocity deviation $v$ and the acceleration/braking time $\tau$
in the form of the equalities (\ref{a}).
As a result, the dependencies of $\tau$ and $v$ on the headway deviation
$\eta$ are given by
\begin{equation}
\tau = \frac{\tau_{0}}{1+\eta^{2}},\qquad
v=\frac{\tau_{0}\eta}{1+\eta^{2}}. \label{9} \end{equation} Note
that, under $\eta$ is in the physically meaningful range between $0$
and $1$, the acceleration/braking time is a monotonically
decreasing function of $\eta$,
whereas the velocity deviation $v$ increases with $\eta$ (at $\eta>1$
we have ${\rm d}v/{\rm d}\eta<0$ that has no physical meaning).

Substituting second equality (\ref{9}) into Eq.~(\ref{8a}) yields the
Landau--Khalatnikov relation:
\begin{equation}
\dot\eta=-\frac{\partial \Phi}{\partial \eta} \label{10}
\end{equation}
with the effective potential given by
\begin{equation}
\Phi=\frac{1}{2}\eta^{2}-
\frac{1}{2}\tau_{0}\ln{\left(1+\eta^{2}\right)}.  \label{11}
\end{equation}
For $\tau_{0}<1$, the $\eta$--dependence of $\Phi$ is
monotonically increasing and the only stationary value of $\eta$
equals zero, $\eta_e=0$, so that there is no headway deviations in
this case. If the parameter $\tau_{0}$ exceeds the critical value,
$\tau_{c}=1$, the effective potential assumes the minimum with
non--zero steady state headway deviation $\eta_{e}=\sqrt{\tau_{0}-1}$
and the acceleration/braking time $\tau_{e}=1$.

The above scenario represents supercritical regime of the traffic jam
formation and corresponds to the second--order phase transition.  The
latter can be easily seen from the expansion of the effective
potential (\ref{11}) in power series of $\eta^{2}\ll 1$:
\begin{equation}
\Phi\approx \frac{1-\tau_{0}}{2}\eta^{2}+
\frac{\tau_{0}}{4}\eta^{4}.
\label{11a}
\end{equation}
So the critical exponents are identical to those obtained within the
framework of the mean-field theory \cite{4a}.

The drawback of the outlined approach is that it fails to account for
the subcritical regime of the self--organization that is the reason
for the appearance of traffic jam and analogous to the
first--order phase transition, rather than the second--order one.
So one has to modify the above theory by taking the assumption that
the effective relaxation time $t_{\eta}(\eta)$
increases with headway deviation $\eta$ from initial value
$t_{\eta}/(1+m)$ fixed by a parameter $m>0$
to the final one $t_{\eta}$ \cite{5a}. The simplest
two--parameter approximation is as follows:
\begin{equation}
\frac{t_{\eta}}{t_{\eta}(\eta)}=1+\frac{m}{1+(\eta/\eta_{0})^{2}},
\label{12} \end{equation}
where $0<\eta_{0}<1$. The expression for the effective potential
(\ref{11}) then changes by adding the term
\begin{equation}
\Delta \Phi=\frac{m}{2}\eta_{0}^{2}
~\ln\left(1+\frac{\eta^{2}}{\eta_{0}^{2}}\right)
\label{13} \end{equation}
and the stationary values of $\eta$ are
\begin{eqnarray}
\eta_{e}^{m}&=&\eta_{00}
\left\{
1\mp \left[
1+\eta_{0}^{2}\eta_{00}^{-4}(\tau_{0}-\tau_{c})
\right]^{1/2}
\right\}^{1/2}, \label{14}\\
2\eta_{00}^{2}&\equiv& (\tau_{0}-1)-\tau_{c}\eta_{0}^{2},\qquad
\tau_{c}\equiv 1+m.\nonumber
\end{eqnarray}
The upper sign in the right-hand side of Eq.~(\ref{14}) is for the
value at the unstable state $\eta^{m}$ where the effective potential
$\Phi+\Delta \Phi$ has the maximum, the lower one corresponds to the stable
state $\eta_{e}$.
The corresponding value of the stationary
acceleration/braking time
\begin{equation}
\tau^{m}=\frac{1+\eta_{00}^{2}+\sqrt{\left(1+\eta_{00}^{2}\right)^{2}-
\left(1-\eta_{0}^{2}\right)\tau_{0}}}{1-\eta_{0}^{2}}
\label{15}
\end{equation}
smoothly increases from the value
\begin{equation}
\tau_{\rm m}=1+\eta_{0}~\sqrt{m\over 1-\eta_{0}^{2}}
\label{16}
\end{equation}
at the parameter $\tau_{0}=\tau_{c0}$ with
\begin{equation}
\tau_{c0}=\left(1-\eta_{0}^{2}\right)\tau_{\rm m}^{2}
\label{17}
\end{equation}
to the marginal value $\tau_c=1+m$ at $\tau_{0}=\tau_c$.

\section{Results}

The $\tau_{0}$--dependencies of $\eta_{e},\,\eta^{m}$, and $\tau_{e}$ are
depicted in Fig.~1. As is shown in Fig.~1a, under the adiabatic
condition $t_{\tau}, t_{v}\ll t_{\eta}$ is met and the parameter $\tau_{0}$
slowly increases being below $\tau_{c}$, no
traffic jam can form. At the point $\tau_{0}=\tau_{c}$ the stationary headway
deviation $\eta_{e}$ jumps upward to the value $\sqrt{2}\eta_{00}$
and its further smooth increase is determined by Eq.~(\ref{14}). If
the parameter $\tau_{0}$ then goes downward, the headway deviation $\eta_{e}$
continuously decreases up to the point, where
$\tau_{0}=\tau_{c0}$ and $\eta_{e}=\eta_{00}$. At this point
the headway deviation jump-like goes down to zero. Referring to Fig.~1b, the
stationary acceleration/braking time $\tau_e$ shows a linear
increase from $0$ to $\tau_{c}$ with the parameter $\tau_{0}$ being
in the same interval. Then, after the jump down to the value
$(1-\eta_{0}^{2})^{-1}$ at $\tau_{0}=\tau_{c}$, the stationary time
$\tau_{e}$ smoothly
decays to $1$ at $\tau_{0}\gg \tau_{c}$. Under the parameter
$\tau_{0}$ then decreases from above $\tau_{c}$ down to $\tau_{c0}$
the acceleration/braking time $\tau_{e}$ grows.  When the point (\ref{17}) is
reached, the traffic becomes freely moving, so that the stationary
acceleration/braking
time undergoes the jump from the value (\ref{16}) up to the
one defined by Eq.~(\ref{17}). For $\tau_{0}<\tau_{c0}$ again the
parameter $\tau_{e}$ does not differ from $\tau_{0}$. Note that this
subcritical regime is realized provided the parameter $m$, that
enters the dispersion law (\ref{12}), is greater than value
\begin{equation}
m_{\rm min} = \frac{\eta_{0}^{2}}{1-\eta_{0}^{2}}.
\label{18} \end{equation}

Clearly, according to the picture described, the jamming generation
is characterized by the well pronounced hysteresis: the cars
initially being at motion with optimal headway between them, begin to
deviate only if the acceleration/braking time $\tau_0$ of cars
exceeds its limiting value $\tau_{c}=1+m$, whereas the
acceleration/braking time $\tau_{c0}$ needed for uniform car flow is
less than $\tau_{c}$ (see Eqs.~(\ref{16}), (\ref{17})). This is the
case in the limit $t_{\tau}/t_{\eta}\to 0$ and the hysteresis loop
shrinks with the growth of the adiabaticity parameter $\delta\equiv
t_{\tau}/t_{\eta}$. In addition to the smallness of $\delta$, the
adiabatic approximation implies the ratio
$t_{v}/t_{\eta}\equiv\epsilon$ is also small. In contrast to the
former, the latter does not seem to be realistic for the system under
consideration, where, in general, $t_{v}\approx t_{\eta}$. So it is of
interest to study to what extent the finite value of $\epsilon$ could
change the results.

Owing to the condition $\delta\ll 1$, Eq.~(\ref{8c}) is still
algebraic and $\tau$ can be expressed in terms of $\eta$ and $v$. As
a result, we derive the system of two nonlinear differential
equations that can be studied by the phase portrait method \cite{5a}. The
phase portraits for various values of $\epsilon$ are displayed in
Fig.~2, where the center $O$ represents the stationary state and the
saddle point $S$ is related to the maximum of the effective potential. As is
seen from the figure, independently of $\epsilon$, there is the
universal section -- the "mainstream", that attracts
all phase trajectories and its
structure is appeared to be almost insensitive to changes in
$\epsilon$. Analysis of time dependencies $v(t)$ and $\eta (t)$
reveals that the headway and velocity deviations slow down
appreciably on this section in comparison to the rest parts of
trajectories that are almost rectilinear (it is not difficult to see
that this effect is caused by the smallness of $\delta$). Since the
most of time the system is in vicinity of the "mainstream", we
arrive at the conclusion that finite values of $\epsilon$ do not
affect qualitatively the above results obtained in the adiabatic
approximation.

\section{Discussion}

According to the above consideration, the simplest picture
of the dissipative dynamic
of traffic flow in a homogeneous car-following model can be represented
within the framework of Lorenz model,
where the headway $\eta$ and velocity $v$ deviations
play a role of an order parameter and
its conjugate field, respectively,
and the acceleration/braking time $\tau$ is a control parameter. The
model is examined to show that a jam creates if the car
characteristic $\tau_0$ is larger than the critical
magnitude $\tau_c$. The above pointed out dissipative regime
is inherent in the systems with small values of the relaxation time
$t_\tau$ for acceleration/braking, being apparently a characteristic
of a car-driver, and large ones $t_\eta$, $t_v$
for the headway and velocity deviations.
According to Ref. \cite{5a}, in the opposite case $t_\tau\ge t_\eta,~t_v$,
the system behaves in auto-oscillation or stochastic manners.

It is worth-while to note that the above synergetic scheme allows us
to explain the collective phenomena of jamming
transition in the $N$-body problem with $N\to \infty$. Then, the question
arises: why exactly three variables (the headway and velocity deviations
$\eta, v$, and acceleration/braking time $\tau$) permit to explain the
nontrivial behaviour of $N$-body problem? The answer to this question gives
the theorem by Ruelle and Takens: nontrivial collective
behaviour of many-body system (type of the strange attractor) can be
represented only in the case, when
number of variables is not less than three \cite{3a}. The interpretation
of this fact is the simplest: the first of the freedom degrees can
be chosen as the way along the phase trajectory, the second one
corresponds to the negative Lyapunov exponent, ensuring an attraction
to this trajectory, the third one acts in opposite manner to give
repulsion. In our case of the self-organization process, the second $v$ and
third $\tau$ freedom degrees provide the positive and negative feedbacks
in Eqs. (\ref{3}), (\ref{6}).

The last question to our approach is why the just Lorenz scheme allows us to
describes main peculiarities of the jamming transition?
The answer is that this is the simplest approach, permitting to catch on
the self-organization effects, just as the Landau phenomenological theory of
phase transitions describes the great variety of thermodynamical
transformations in the simplest way \cite{6a}.
Let us note in this connection that the effective potential given by
the sum of equalities (\ref{11}), (\ref{13}) acts the part of the
Landau free energy. But the above stated synergetic scheme has
the principle difference
from the Landau-type theory \cite{2a} because the former takes
into account a feedback of thermostat
(the velocity deviation and the acceleration/braking time)
with subsystem under consideration (the headway deviation),
whereas the latter does not.

\section*{Acknowledgment}

We are grateful to anonymous reviewer for constructive criticism
that promotes to improve the above scheme exposition.

\vspace*{-0.8cm}

%\begin{references}

\newpage
\begin{center}
{\bf CAPTIONS}
\end{center}

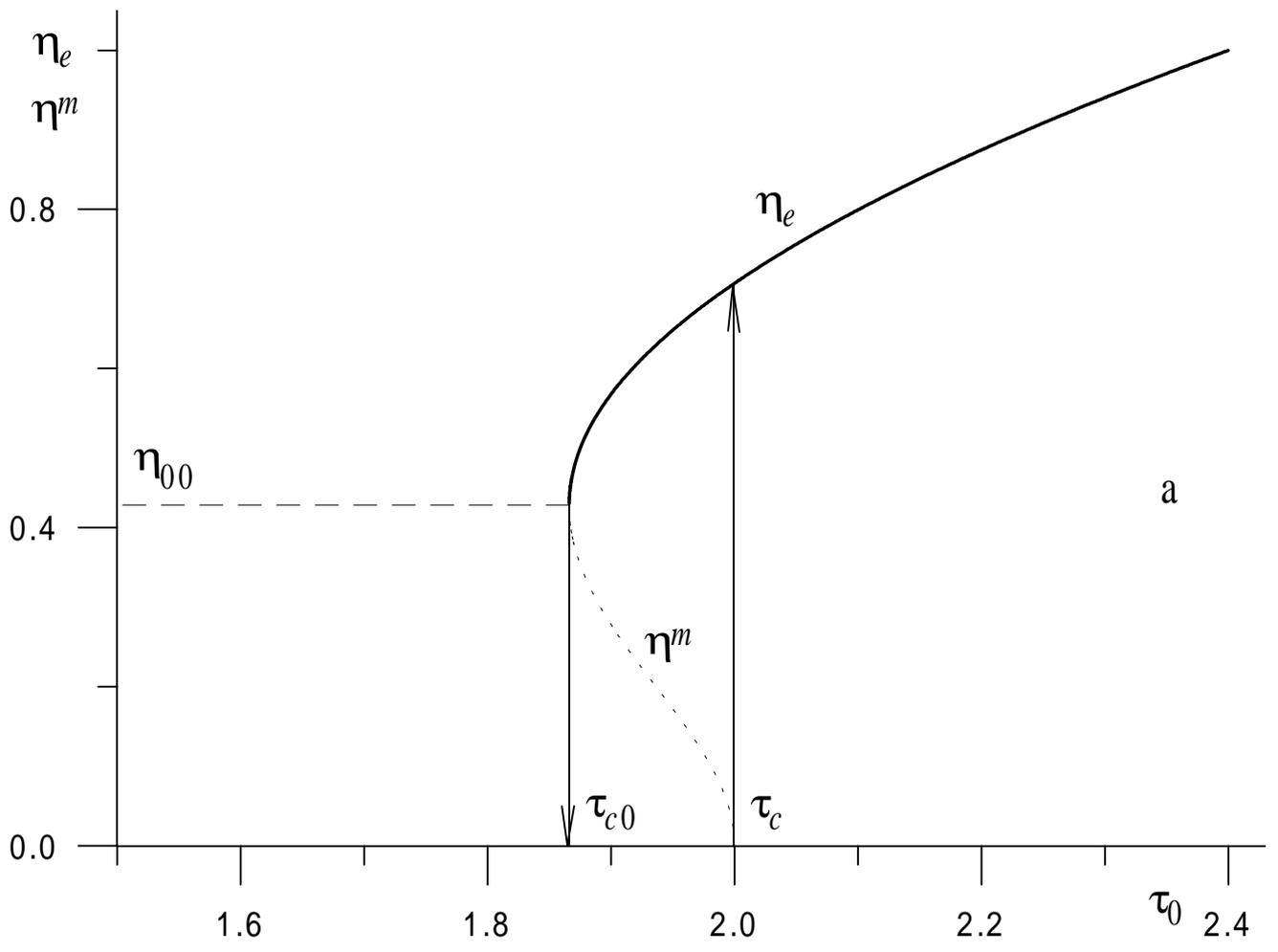
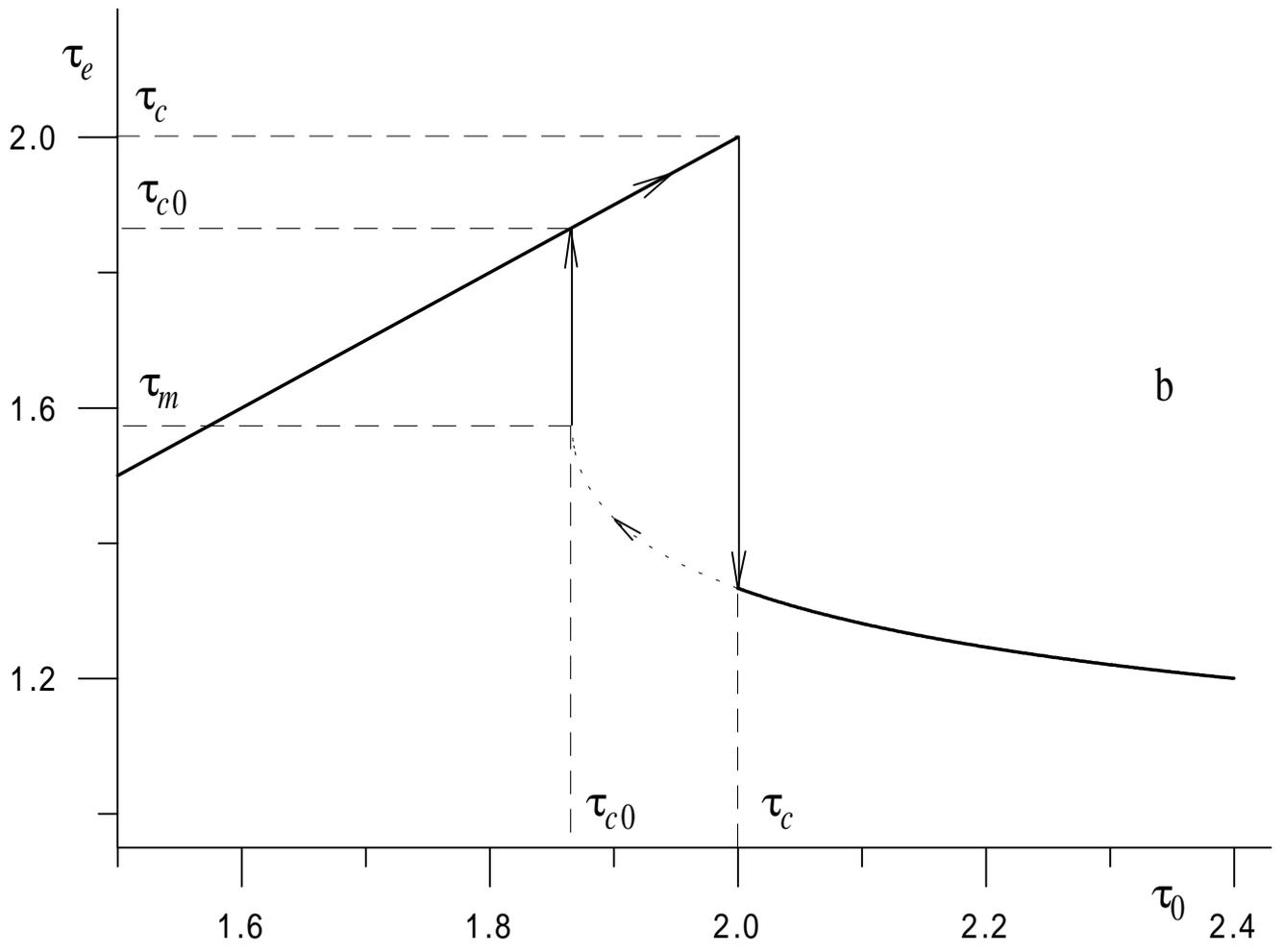
Fig.~1. The $\tau_{0}$--dependencies of the stationary values of:
(a) headway deviations $\eta_{e}$, $\eta^{m}$;
(b) acceleration/braking time $\tau_{e}$.
The arrows indicate the hysteresis loop.

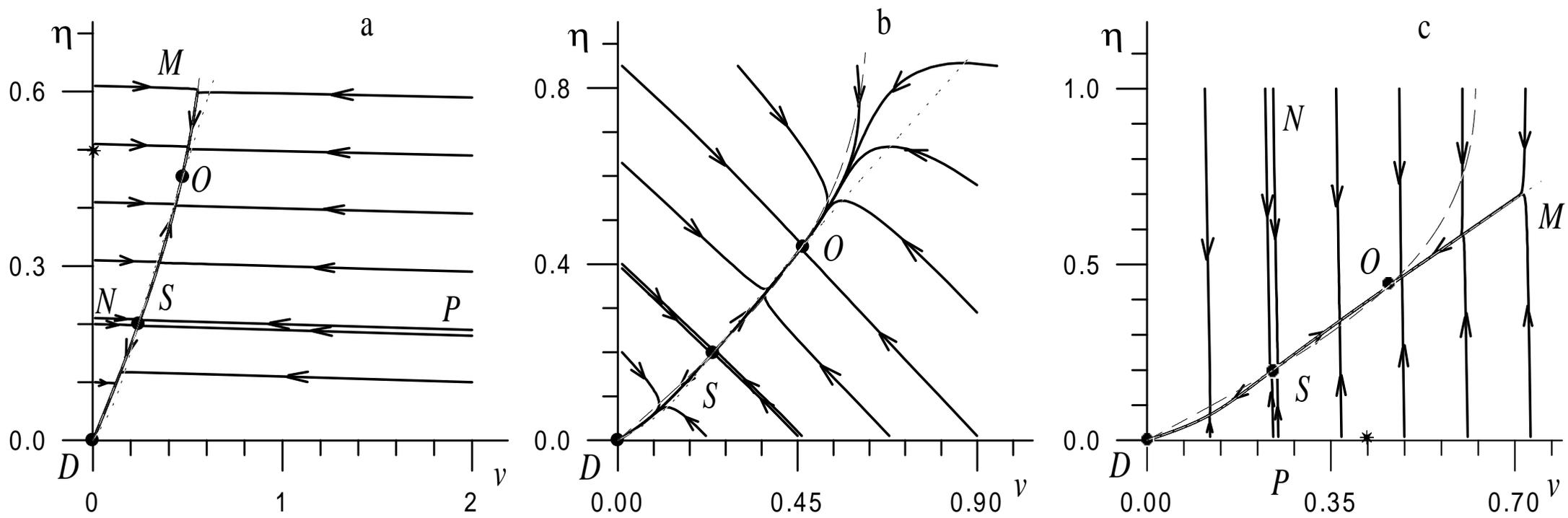
Fig.~2. Phase portraits in the $\eta-v$ plane at $m=1, ~\eta_0 =
0.1, ~\tau_{0}=1.25\tau_{c}$ for: (a) $\epsilon=10^{-2}$;
(b) $\epsilon=1$; (c) $\epsilon=10^{2}$.

\end{document}